%% file: AGILE_first_catalog_revised_v2.tex
\documentclass{aa}

\usepackage{txfonts}
\usepackage{natbib}
\usepackage{epsfig}
\usepackage{lscape}

\usepackage{rotating}

\bibpunct{(}{)}{;}{a}{}{,} 

\newcommand{\Mev}{{\rm MeV~}}

\newcommand{\beq}{\begin{equation}}
\newcommand{\eeq}{\end{equation}}

\newcommand{\cp}{}
\newcommand{\new}{}
\newcommand{\nnew}{}

\usepackage{graphicx}
\graphicspath{{converted_graphics/}}
\begin{document}

\title{First AGILE Catalog of High-Confidence Gamma-Ray Sources }         

\author{C. Pittori$^{1}$, F. ~Verrecchia$^{1}$, A.W.~Chen$^{2,3}$, A.~Bulgarelli$^{4}$,
A.~Pellizzoni$^{5}$,  A.~Giuliani$^{2,3}$, S.~Vercellone$^{6}$, F.~Longo$^{7,8}$,
M. Tavani$^{9,10,11,3}$, P.~Giommi$^{1,12}$,
G.~Barbiellini$^{7,8,3}$,
M.~Trifoglio$^{4}$,
F. ~Gianotti$^{4}$,
A.~Argan$^{9}$,
A.~Antonelli$^{13}$,
F.~Boffelli$^{14}$,
P.~Caraveo$^{2}$,
P.~W.~Cattaneo$^{14}$,
V.~Cocco$^{10}$,
S.~Colafrancesco$^{1,12}$,
{\new T. Contessi$^{2}$,}
E.~Costa$^{9}$,
S.~Cutini$^{1}$,
F.~D'Ammando$^{9,10}$,
E.~Del~Monte$^{9}$,
G.~De~Paris$^{9}$,
G.~Di~Cocco$^{4}$,
G.~Di~Persio$^{9}$,
I.~Donnarumma$^{9}$,
Y.~Evangelista$^{9}$,
G.~Fanari$^{1}$,
M.~Feroci$^{9}$,
A.~Ferrari$^{3,15}$,
M.~Fiorini$^{2}$,
F.~Fornari$^{2}$,
F.~Fuschino$^{4}$,
T.~Froysland$^{3,11}$,
M.~Frutti$^{9}$,
M.~Galli$^{16}$,
D.~Gasparrini$^{1}$,
C.~Labanti$^{4}$,
I.~Lapshov$^{9,17}$,
F.~Lazzarotto$^{9}$,
F.~Liello$^{9}$,
P.~Lipari$^{18,19}$,
E.~Mattaini$^{2}$,
M.~Marisaldi$^{4}$,
{\new M.~Mastropietro$^{9,21}$, }
A.~Mauri$^{4}$,
F.~Mauri$^{14}$,
S.~Mereghetti$^{2}$,
E.~Morelli$^{4}$,
E.~Moretti$^{7,8}$,
A.~Morselli$^{11}$,
L.~Pacciani$^{9}$,
F.~Perotti$^{2}$,
G.~Piano$^{9,10,11}$,
P.~Picozza$^{10,11}$,
{\new M.~Pilia$^{22,2,5}$,}
C.~Pontoni$^{3,8}$,
G.~Porrovecchio$^{9}$,
B.~Preger$^{1}$,
M.~Prest$^{8,22}$,
R.~Primavera$^{1}$,
G.~Pucella$^{9}$,
M.~Rapisarda$^{20}$,
A.~Rappoldi$^{14}$,
E.~Rossi$^{4}$,
A.~Rubini$^{9}$,
S.~Sabatini$^{10}$,
P.~Santolamazza$^{1}$,
{\cp E.~Scalise$^{9}$,}
P.~Soffitta$^{9}$,
S.~Stellato$^{1}$,
{\cp E.~Striani$^{10}$,}
F.~Tamburelli$^{1}$,
A.~Traci$^{4}$,
A.~Trois$^{9}$,
E.~Vallazza$^{8}$,
V.~Vittorini$^{9,3}$,
A.~Zambra$^{2,3}$,
D.~Zanello$^{18,19}$,
and L.~Salotti$^{12}$  }

\institute{ $^{1}$  ASI Science Data Center, ESRIN, I-00044 Frascati (RM), Italy.\\
$^{2}$ INAF-IASF Milano, via E. Bassini 15, I-20133 Milano, Italy.\\
$^{3}$ Consorzio Interuniversitario Fisica Spaziale (CIFS), villa Gualino - v.le Settimio Severo 63, I-10133 Torino, Italy.\\
$^{4}$ INAF-IASF Bologna, via Gobetti 101, I-40129 Bologna, Italy.\\
$^{5}$ Osservatorio Astronomico di Cagliari, loc. Poggio dei Pini, strada 54, I-09012, Capoterra (CA), Italy. \\
$^{6}$ INAF-IASF Palermo, Via Ugo La Malfa 153, I-90146 Palermo, Italy. \\
$^{7}$ Dip. Fisica, Universit\`a di Trieste, via A. Valerio 2, I-34127 Trieste, Italy.\\
$^{8}$ INFN Trieste, Padriciano 99, I-34012 Trieste, Italy.\\
$^{9}$ INAF-IASF Roma, via del Fosso del Cavaliere 100, I-00133 Roma, Italy.\\
$^{10}$ Dipartimento di Fisica, Universit\`a Tor Vergata, via della Ricerca Scientifica 1,I-00133 Roma, Italy.\\
$^{11}$ INFN Roma Tor Vergata, via della Ricerca Scientifica 1, I-00133 Roma, Italy.\\
$^{12}$  Agenzia Spaziale Italiana, viale Liegi 26, I-00198 Roma, Italy.\\
$^{13}$ Osservatorio Astronomico di Roma, Via di Frascati 33, I-00040 Monte Porzio Catone, Italy. \\
$^{14}$ INFN Pavia, via Bassi 6, I-27100 Pavia, Italy.\\
$^{15}$ Dipartimento di Fisica, Universit\`a di Torino, via P. Giuria 15, I-10126 Torino, Italy.\\
$^{16}$ ENEA Bologna, via don Fiammelli 2, I-40128 Bologna, Italy.\\
$^{17}$ IKI, Moscow, Russia.\\
$^{18}$ INFN Roma 1, p.le Aldo Moro 2, I-00185 Roma, Italy.\\
$^{19}$ Dip. Fisica, Universit\`a La Sapienza,p.le Aldo Moro 2, I-00185 Roma, Italy.\\
$^{20}$ ENEA Frascati, via Enrico Fermi 45, I-00044 Frascati (RM), Italy. \\
$^{21}$ CNR, IMIP, Area Ricerca Montelibretti (RM), Italy.\\
$^{22}$ Dip. Fisica, Universit\`a dell'Insubria, Via Valleggio 11, I-22100 Como, Italy.
}

\offprints{C. Pittori, \email{pittori@asdc.asi.it} }
\date{received; accepted}
\authorrunning {C. Pittori et al.}
\titlerunning {First AGILE Catalog}

\abstract{We present the first catalog of high-confidence
$\gamma$-ray sources detected by the AGILE satellite during observations performed
from July 9, 2007 to June 30, 2008. Catalogued sources are detected
by merging all the available data over the entire time period.
AGILE, launched in April 2007, is an ASI mission devoted to
$\gamma$-ray observations in the 30 MeV -- 50 GeV energy range, with
simultaneous X-ray imaging capability in the 18--60 keV band. This
catalog is based on Gamma-Ray Imaging Detector (GRID) data for
energies greater than 100 MeV. For the first AGILE catalog we adopted
a conservative analysis,
with a high-quality  event filter 
optimized to select $\gamma$-ray events within the central zone of the
instrument Field of View (radius of 40$^{\circ}$).
 {\cp This is a significance-limited (4 $\sigma$) catalog,
and it is not a complete flux-limited sample due to the
non-uniform first year AGILE sky coverage.}
The catalog includes {\cp 47 sources},
21 of which are associated
with confirmed or candidate pulsars,
13 with Blazars (7 FSRQ, 4 BL Lacs, 2 unknown type), 2 with
HMXRBs, 2 with SNRs, 1 with a colliding-wind binary system,
8 with unidentified sources.}
\vskip 2 truecm

\maketitle

\section{Introduction}
AGILE (Astrorivelatore Gamma ad Immagini LEggero) \cite{Tavani,Tavani1}
is a mission of the Italian Space Agency (ASI)
devoted to $\gamma$-ray astrophysics in the 30 MeV -- 50 GeV, and 18 -- 60 KeV energy ranges.
{\new AGILE was successfully launched on April 23, 2007 in a {$\sim$}550 km 
equatorial orbit with low inclination angle, {$\sim$} 2.5$^{\circ}$.
}
{\cp High energy $\gamma$-ray astrophysics is entering a new
challenging phase of discovery.
During the 1970's and 1980's, the SAS-2 ~\cite{SAS-2} and COS-B
~\cite{COS-B1,COS-B2} space missions 
discovered the very first cosmic
$\gamma$-ray sources around 100 MeV, but
our knowledge of high energy cosmic $\gamma$-ray emission and
phenomena up to now was mainly based on the remarkable results
obtained by the EGRET instrument, on board of the Compton
Gamma-Ray Observatory (CGRO) \cite{egret}. Nearly 300 $\gamma$-ray
sources above 100 \Mev were detected by EGRET \cite{1999ApJS..123...79H} during the period
from April 22, 1991 to October 3, 1995; however, only a small
fraction of them ($\sim 30 \% $) is currently identified. 
Many sources are variable or transient on short timescales, and our
understanding of many high energy phenomena is still preliminary.

AGILE is the first $\gamma$-ray mission operating
in space after almost ten years since the end of EGRET operations.
AGILE was the only mission entirely dedicated to high
energy astrophysics above 30 \Mev during the period
April 2007- June 2008.
It is currently operating together with the Fermi Gamma-Ray Space Telescope
(formerly GLAST), launched on June 11, 2008 \cite{GLAST,LAT}.
The highly innovative AGILE instrument is the first of a
new generation of high-energy space missions
based on 
solid-state silicon technology,
expected to substantially advance our knowledge
in several research areas including the study of
Active Galactic Nuclei, Gamma-Ray Bursts, pulsars, unidentified
$\gamma$-ray sources, Galactic compact objects, supernova remnants, etc.}
\par
{\cp 
The AGILE Payload detector consists of
the Silicon Tracker (ST) \citep{2001AIPC..587..754B,2003NIMPA.501..280P},
the X-ray detector SuperAGILE \citep{2007NIMPA.581..728F},
the CsI(Tl) Mini-Calorimeter (MCAL) \citep{2006SPIE.6266E.110L},
and an anticoincidence system (ACS) \citep{2006NIMPA.556..228P}.
The combination of ST, MCAL and ACS forms the Gamma-Ray Imaging
Detector (GRID).
Accurate timing, positional and attitude information is provided by the 
Precise Positioning System and the two Star Sensors units.}
The Silicon Tracker,
based on the process of photon conversion into electron-positron pairs,
is the core of the AGILE-GRID.
It consists of a total of 12
trays, the first 10 of which are capable of converting $\gamma$-rays by a
Tungsten layer {\cp tracked by silicon microstrip detectors 
providing
the two orthogonal coordinates for each element (point)
along the track.}
AGILE-GRID event processing is operated by on-board trigger logic algorithms
\citep{PDHU}
and by {\cp on-ground  event filtering.  In this paper 
we use the on-ground GRID
event filter called ``F4"  \cite{giulianik1}.} 

{\new
During its first year in orbit AGILE surveyed
the $\gamma$-ray sky and detected many galactic and extragalactic sources.
The AGILE Commissioning phase ended on July 9, 2007, and the following
Science Verification Phase lasted about four months, up to November 30, 2007.
On December 1, 2007 the baseline nominal observations and pointing plan of Cycle-1 started
together with the Guest Observer program.}
In this paper we present the first catalog of high-confidence $\gamma$-ray sources detected by
AGILE including data from July 9, 2007 to June 30, 2008, thus covering
Science Verification Phase data and the first seven months of the Cycle-1.
Catalogued sources are detected by merging all the available data over the entire time period.

{\cp   
This is a significance-limited catalog including only sources above 4 $\sigma$
extracted from the sample of AGILE detections obtained
with a conservative data analysis, as described in the following sections.
The catalog sensitivity is nonuniform
reflecting the inhomogeneous first year AGILE sky coverage.
{\new The first year exposure (see Fig. \ref{fig:expo})
was focused mainly towards the Galactic plane,
mostly in the Carina-Crux and in the Cygnus regions.
The average effective AGILE-GRID exposure time 
across the sky, for the chosen F4 event filter,
is $< T_{exp}> \simeq 0.8 \times 10^6$ s, with peak values of 
$\sim 7 \times 10^6 $ s.}
{\cp For a given statistical significance the limiting 
point-source flux varies with position, owing to
the diffuse $\gamma$-ray emission which represents a nonuniform
background over which the point-like sources are seen.
The Galactic diffuse continuum $\gamma$-ray emission dominates other components and has a
wide distribution with most emission coming from the Galactic plane.
We detect limiting fluxes of about 2 $\times 10^{-7}$ ph cm$^{-2}$
s$^{-1}$ with $\sim 4 \sigma$ statistical significance at galactic
latitudes $|b|>10^{\circ}$.}

{\new The outline of the paper is as follows: in Sect. \ref{sec:response} we briefly discuss
the AGILE-GRID response characteristics and in Sect.
\ref{sec:diffuse} we describe the AGILE diffuse $\gamma$-ray model
used in the data analysis. 
In Sect.\ref{sec:dataflow} we describe the satellite pointing strategy and
data flow. We then present in Sect. \ref{sec:method} the data reduction
and analysis method used to build the first AGILE catalog. Our
results and the list of detected 
high-confidence $\gamma$-ray
sources are shown in Sect. \ref{sec:list}. Finally in Sect.
\ref{sec:concl} we discuss our results and make some concluding
remarks.}
} 

{\cp
\section{AGILE-GRID Response Characteristics}
\label{sec:response}

{\new The AGILE-GRID in-flight calibration 
during the first year and a half of observations
has been recently completed 
and details of the instrument response characteristics
will be given in Tavani et al. 2009b. 
The results are consistent with pre-launch simulations and instrument tests
\cite{Tavani1}.}

The energy-dependent in-flight GRID instrument point spread function (PSF)
has a FWHM (Full-Width at Half-Maximum)
of approximately 3.5$^{\circ}$ at 100 MeV, 
and gradually improves at higher energies.
AGILE PSF is better than that of EGRET by a factor
of $\sim 2$ above 400 {\rm MeV}.
The GRID effective area, as determined by in-flight calibrations, 
reaches 500 $\rm cm^2$ at several hundred MeV, depending on the GRID
event filter used.
The conservative event filter F4, chosen for our analysis,
applies tight event selection cuts to eliminate an higher fraction of 
possible particle background counts. This filter is
optimized to select $\gamma$-ray events within the central 
Field of View (FoV) zone ($\sim 40^{\circ}$ radius)
at the expenses of the effective area.
 In the energy range
200 -- 400 MeV and at $\sim 30^{\circ}$ off-axis, 
the average effective area for the F4 filter is
$<A_{eff}>_{(F4)} \sim 300$ cm$^2$ . 
{\new It can be parametrized as a function of the off-axis angle 
$\theta$, in the range $\in[0 \div \theta_M]$, as:}
\beq
<A_{eff}(\theta)>_{(F4)}= A_0~[1-(\frac{\theta}{\theta_M})^{\alpha_1}]^{\alpha_2}
\eeq
where: $A_0=366$ cm$^2$, $\theta_M=64^\circ$, $\alpha_1=3$ and $\alpha_2=2$.

Both AGILE PSF and effective area
are characterized by a very good off-axis performance
and are well calibrated up to almost $60^\circ$,
showing very smooth variations with the angle relative to the
instrument axis.
}


\section{AGILE Diffuse Gamma-ray Model}
\label{sec:diffuse}

{\cp  
{\new In the data analysis
we use the AGILE diffuse emission model \cite{giuliani04,giulianidiff}
for diffuse $\gamma$-ray background counts predictions.
Diffuse $\gamma$-ray emission includes a combination of two components:
(1) diffuse emission from the Galactic interstellar medium
and (2) an approximately isotropic extragalactic component, 
plus possible contributions from unresolved and faint point sources.
Diffuse emission coming from the Galactic plane dominates other components
and, as in the EGRET model \citep{1997ApJ...481..205H}, 
it is assumed to
be produced by the interaction of cosmic rays with the
interstellar medium through three physical processes:
proton-proton collision, Bremsstrahlung and inverse Compton
emission.}

The AGILE diffuse emission model substantially improves  the
previous EGRET model by using state-of-the-art neutral hydrogen (HI) and CO
updated maps in order to
model the matter distribution in the Galaxy.
It is based on a 3-D grid with $0.25\times 0.25$
square degrees binning in Galactic longitude and latitude, and a
0.2 kpc step in distance along the line of sight. 
Concerning the distribution of neutral hydrogen, we used the
Leiden-Argentine-Bonn (LAB) Survey of Galactic HI
\cite{kalberla05}. The LAB survey improves previous results
especially in terms of sensitivity (by an order of magnitude),
velocity range and resolution. 
In order to properly project the velocity-resolved radio data, we
used the Galactic rotation curves parameterized by Clemens et al. 1985.
The  detailed and relatively
high-resolution distribution of molecular hydrogen is obtained
from the CO observations described in Dame et al. 2001. The CO
is assumed to be a tracer of molecular hydrogen, through a known
ratio between hydrogen density and CO radio emissivity.
Cosmic rays can emit $\gamma$-rays through the inverse Compton
mechanism due to their interaction with photons of the
cosmological background and of the interstellar radiation
field (ISRF). In order to account for this component we use the
analytical model proposed by Chi \& Wolfendale 1991. It
describes the ISRF as the result of three main contributions: far
infrared (due to dust emission), near infrared, and optical/UV
(due to stellar emission).
{\new The distribution of cosmic rays (both protons and electrons)
in the Galaxy is 
obtained using the GALPROP
cosmic-ray model  (Strong. et al., 2004 and Strong 2007). 
}
} 

\section{AGILE Data Flow and Cycle-1 Observational Program}
\label{sec:dataflow}


{\cp AGILE satellite raw Telemetry data are down-linked approximately every 100 minutes to the ASI Malindi
ground station in Kenya and transmitted first to the
Mission Control Center at Telespazio, Fucino, and subsequently to the AGILE Data Center (ADC)
for data reduction, scientific processing and archiving.}
The ADC is the scientific component of the
AGILE ground segment and is part of the ASI Science Data Center
(ASDC) located in Frascati (Italy).
The ADC includes scientific personnel from both the ASDC and the AGILE Team.
More details on the ADC organization and tasks will be given in 
Pittori et al. 2009.


The AGILE pointings are subject to strict constraints requiring that the fixed solar
panels {\cp always be} oriented within 3$^{\circ}$ from the Sun direction.
{\cp AGILE pointings are called Observation Blocks (OBs) and usually consist of predefined
long exposures, typically lasting 10 -- 30 days, drifting
about $1$ degree per day
with respect to the initial boresight direction to match solar panels
illumination constraints.}
The large GRID Field of View ($\sim 2.5$ sr)
and the low altitude orbit imply that, for most pointing directions,
the Earth (partially) occults the field of view,
{\cp  
thus the observing efficiency and
exposure for a given source varies depending on its coordinates.
In order to eliminate the Earth albedo $\gamma$-ray contamination
originating from interactions of cosmic rays with
the upper atmosphere, a limb-angle cut was applied
for all the $\gamma$-ray events with reconstructed directions
smaller than $80^{\circ}$ with respect to the satellite-Earth vector. 
With this event selection 
we do not {\nnew expect} 
systematic effects caused by albedo photon background fluctuations.}

A predefined AGILE Baseline Pointing Plan, aimed at reaching specific scientific
goals which maximize the scientific output of the mission, is
{\new made public in advance at the AGILE web pages at ASDC\footnote{{\it http://agile.asdc.asi.it}} 
to allow for}  the organization of
multi-wavelength campaigns.
Part of the AGILE Science Program is open to Guest Observers 
on a competitive basis
through Announcements of Opportunity. 
Guest Observers can apply for data {\cp  which will be} collected
within the Pointing Plan. In case of Target of Opportunity (ToO)
observations, the baseline Plan is {\cp interrupted and resumed
 at the end of the ToO, so that usually a ToO
replaces some of the foreseen baseline pointings, and does not
shift in time the execution of the remaining  planned observations.}
{\cp
The ADC web pages 
provide interactive tables for both the predefined AGILE Baseline Pointing Plan
and the actual list of pointings, including previously unforeseen ToOs.}
 
In this paper we analyzed AGILE-GRID data of the 63 Observation Blocks
reported in Table \ref{table:pointings},
covering the period from July 9, 2007 to June 30, 2008.
{\cp
The total $\gamma$-ray exposure and counts maps 
obtained over the selected period with the F4 filter, 
in Aitoff projection and Galactic coordinates,
are shown in Fig.\ref{fig:expo} and Fig.\ref{fig:counts}, respectively.
}



\section{AGILE Data Reduction and Analysis}
\label{sec:method}

Raw AGILE telemetry received at ADC is archived and transformed in  FITS format through the
AGILE Pre-Processing System (TMPPS) \cite{trifoglio}.
All GRID data are then routinely processed 
using the scientific
data reduction software tasks developed by the AGILE
instrument team and integrated into an automatic 
pipeline system developed at the ASI Science Data Center.
The first step of the 
pipeline converts on a contact-by-contact basis
{\new the satellite 
data time into Terrestrial Time (TT), and performs some
preliminary calculations and {\cp unit} conversions.}
A second step consists of the $\gamma$-ray event selection. We use an
AGILE-GRID specific implementation of the Kalman Filter technique
for track identification, event direction and energy 
reconstruction \cite{giulianik1,arem}. 
{\cp 
A quality flag is assigned to each event
depending on whether it is recognized as a $\gamma$-ray event, a
charged particle, a ``single-track" event, or an event of
{\cp uncertain classification. } An AGILE auxiliary (LOG) file is
created, containing all the spacecraft information relevant
to the computation of the effective exposure and live-time.
Finally, the event directions in sky coordinates are reconstructed
and reported in  the AGILE event files (EVT), excluding events
flagged as charged background particles.
This step produces the 
Level-2 archive of LOG and EVT files on temporal intervals
of few hours.
A third step of the pipeline 
creates quick look (QL) counts, exposure, and {\cp diffuse
$\gamma$-ray emission maps } on different  time scales: 
days, weeks and 
daily increments during the OB timescale.

At the completion of each OB, we run the AGILE Standard Analysis
OB pipeline which removes  the data corresponding to
repointing slews and occasional losses of fine-pointing attitude.
{\new GRID data used in our analysis have been processed with the 
 standard software and in-flight calibrations available at
the time of writing}\footnote{ Software build version: BUILD
GRID\_STD\_16 and BUILD GRID\_SCI\_15.2}.
We used the high-quality
F4 event filter\footnote{New filter algorithms of large efficiency
and  optimized over a wider FoV have been developed and 
will be distributed by ADC during  Cycle-2.
}, whose response characteristics were described in Sect.\ref{sec:response}.
} 

The standardized and cleaned OB Level-2 archive is the basis for creating 
Guest Observers data packets and for the data merging used
to build this first catalog.

\subsection{Data Merging from the OB Archive}

In order to merge the data from different observing periods over the whole sky,
we produced sets of FITS images in the ARC projection \citep{2002A&A.395..1077C}
in Galactic coordinates, with a radius of $40^{\circ}$ and a bin size of
$0.25^{\circ} \times 0.25^{\circ}$, oriented with the north Galactic pole facing upward.
{\cp The centers of the maps were chosen according to the HEALPix
(Hierarchical Equal Area isoLatitude Pixelization) algorithm
 \citep{2005ApJ...622..759G}
with $N_{side} = 4$,
for the coverage of the full sky with 192 maps, whose centers are at constant latitude.}
{\cp HEALPix algorithm
produces a subdivision of a spherical surface in which each pixel covers the same surface area as every other pixel.
Note however that the HEALPix projection in FITS \citep{2007MNRAS.381..865C} 
is not used here.
{\new Only
the property of the HEALPix grid that the pixel centers occur on a discrete number of
rings of constant latitude is used
to represent all-sky $\gamma$-ray data binned in sky coordinates.}
The circular sky areas defined
by a centroid and a radius constructed around the 192 HEALPix points
are hereafter called {\cp``}rings".}

For each 12 hour period,
we produced maps of counts and exposure in the full energy band
$E=100$ MeV -- 50 GeV
{\cp in rings yielding at least 20 minutes of effective
exposure time within  $30^{\circ}$ from each
HEALPix point}
\footnote{We describe
here our choice of parameters for maps generation.
To reduce the particle background
contamination, only events tagged as con\-fir\-med
$\gamma$-ray events were selected (filtercode=5). The South Atlantic
Anomaly data were excluded (phasecode=18) and all the
$\gamma$-ray events whose reconstructed directions with respect to
 the satellite-Earth vector is smaller than $80^{\circ}$ (albrad=80) were
also rejected, to eliminate the Earth albedo contamination.}.
The 12-hour maps covering the whole sky were then summed 
over the entire one-year data span
and analyzed with two independent source detection 
algorithms as described in the following Section.

\subsection{Source Detection Method}

The AGILE source detection method is based on a Maximum Likelihood
(ML) analysis to derive, for each candidate source, the best
parameter estimates of source significance, $\gamma$-ray flux, and
source location \cite{chen}.
The ML statistical technique,
already used in the past in the analysis of $\gamma$-ray data
\citep{1996ApJ...461..396M},
compares measured counts in each pixel
with the predicted counts derived from 
the diffuse $\gamma$-ray model
to find statistically significant  excesses 
consistent with the
instrument point spread function. 
{\cp In the analysis we use the AGILE diffuse $\gamma$-ray model 
described in Sect. \ref{sec:diffuse}
for diffuse background (gas) map generation.}
\par
The Likelihood ratio test is then used
to compare the null (diffuse background-only) hypothesis with
{\cp possible presence of point-source components. } 
According to Wilks' theorem (Wilks 1938),
the point source ``Test Statistic" (TS), defined as:
\beq
TS = -2~(\ln L_0 - \ln L_1)
\eeq
is expected to behave as $\chi_1^2$ in the null hypothesis, 
plus terms of order $O(N^{-1/2})$, where N is the number of counts. In practice
for a number of AGILE counts $N>20$,
the significance of a source detection at a certain position
is given by a number of standard deviations $\sigma$ equal to $\sqrt{TS}$.
\par

Our method for source detection consists of three steps:
\begin{itemize}
\item[1)] Preliminary automatic detection of counts map excesses and ML analysis
on the resulting fixed positions. This step was performed with two independent
detection strategies (A and B described below).

\item[2)] Selection of high-confidence detections, according to the criteria
described below.

\item[3)] Refined analysis with a ML multi source task to optimize
source locations and flux estimates.

\end{itemize}

\par

In step 1) two independent automatic source detection strategies {\cp over each ring count map}
were used:
\begin{itemize}
\item[A)]  Identification of possible source locations 
using 
the standard Ximage\footnote{Ximage \citep{1992ASPC...25..100G} 
is part of the NASA's Heasarc Xanadu standard software package for 
multi-mission X-ray astronomy.}
software for astronomical imaging \citep{1992ASPC...25..100G}, 
adapted to $\gamma$-ray data analysis,
and then 
{\cp 
single source ML analysis 
(with the {\it AG\_srctest\_fixed}
task of the AGILE scientific pipeline).} {\cp The
Ximage detection algorithm locates point sources using a sliding-cell method
so that positions and fluxes of each detected source are evaluated in a
box maximizing the signal-to-noise ratio}.

\item[B)] Identification of possible source locations 
using a dedicated algorithm developed by the AGILE Team called SPOT
{\cp based on a wavelet filtering technique adapted to $\gamma$--ray data}
\citep{2008ADASS.XXX...YYY}, 
followed by a multi source ML analysis
(with the task {\it AG\_srclist}) used iteratively.

\end{itemize}

{\nnew
The AGILE SPOT algorithm used as method B) is a two-step procedure 
that extracts the excesses from counts maps
and builds a list of candidate gamma-ray objects 
which are then analyzed by a likelihood method.
To determine counts excesses over the background, the SPOT algorithm
analyzes the binned count maps with a 
smoothing of 1 degree, considers the bins with the largest 
number of counts, and adds to them the neighboring bins, increasing the connected region,
as described in \citep{DiStefano}.
The process ends when another connected region is merged with the first growing region.
At that point, the merging step is reversed and two distinct connected regions are obtained. 
The centroids of all regions obtained in this way 
identify the positions of the gamma-ray candidate sources
to be analyzed by a multi source ML.}
Method A), which uses a single source likelihood analysis,
optimizes detections of isolated $\gamma$-ray sources typically in
extragalactic sky regions, whereas method B) is more efficient in
complex regions such as on the Galactic plane, where multiple
source contributions may contaminate the result. In both cases we
use an analysis radius of $10^{\circ}$ 
and {\cp a single} power-law source model with
spectral photon index $\alpha$. {\cp In our analysis we adopted a
standard value of $\alpha = -2.1$, except for the Vela ($\alpha_{Vela}= -1.69$)
and Geminga ($\alpha_{Gem}= -1.66$)
pulsars. This assumption is motivated by the known Crab-like
spectral properties of the majority of EGRET sources, and by the
relatively small statistical significance of several AGILE
sources, limiting our spectral analysis capability with chosen data sample.
We postpone a detailed spectral analysis of the sources
appearing in this catalog to forthcoming publications.}

We populate two {\cp  databases} with all the results obtained
with the automatic methods A) and B) and in step 2) we
cross-correlate the two sub-samples extracted from the database
with the following conditions:

\begin{itemize}
\item[-] distance of the candidate source location from the center of the ring field of view
has to be less than or equal to $30^{\circ}$,
{\cp in order to perform the $10^{\circ}$ data analysis within the confidence region of the
chosen F4 filter algorithm ($\le 40^{\circ}$). }

\item[-] in each database we create a sub-sample by associating to a
single entry all the detections within a radius of 90'; 

\item[-] for sources appearing in different ring areas, we select only
detections with minimal distance from the center of the ring field of view;

\item[-] we select detections with $\sqrt{(TS)} > 4$, which corresponds to a statistical
significance of about 4 $\sigma$.
\end{itemize}

We obtain:
\begin{itemize}
\item[-] 81 source candidates with source detection method A)
\item[-] 77 source candidates with source detection method B)
\end{itemize}

An initial cross-correlation radius of 90' between the two dataset
was used in order to select  high-confidence galactic and
extragalactic source candidates.

\par
Then in step 3), a manual refined analysis was performed
with a multi source likelihood analysis task, {\it AG\_multi},
to confirm the detection and derive optimized source parameters.
Special care should be used in particular on the galactic plane region ($|b|< 10^{\circ}$)
to deal with possible source confusion and flux contamination.
{\cp In the analysis of complex regions, positioning results
obtained with detection methods A) and B)
were also compared with a third standard peak detection software, SExtractor (Bertin et Arnaud 1996),
adapted to $\gamma$-ray data, using a wavelet filtering and deblending algorithm.
We define ``high-confidence'' those detections which
pass all the requirements described in this Section. 
}


\section{First AGILE Catalog of High-Confidence Gamma-Ray Sources}

\label{sec:list}
{ \cp
The resulting list of validated sources,
detected by using AGILE-GRID data from July 9, 2007 to June 30, 2008 
with the method and criteria described in section \ref{sec:method},
includes 47 high-confidence sources.
The sources of this first AGILE catalog 
are plotted in Fig.~\ref{fig:aitoff}
in galactic sky coordinates, censused in Table \ref{tab:assoc}, including
both confirmed and possible associations, and listed in Table 3. 
}

\begin{table*}[tbh]

\begin{center}

\begin{tabular}{|c c c|}

\hline
& & \\
\multicolumn{1}{|c}{Classification} & 
\multicolumn{1}{c}{Confirmed Counterparts} & 
\multicolumn{1}{c|}{Possible Counterparts} \\[3 pt]

\hline
 & & \\[3 pt]

Pulsar & 7 & 14 \\[3 pt]

Blazar ~~FSRQ type & 4 & 3 \\[3 pt]

Blazar BL Lac type & 4 & -- \\[3 pt]

Blazar Unknown type & -- & 2 \\[3 pt]

CWB & 1 & -- \\[3 pt]

SNR & 2 & -- \\[3 pt]

HMXRB & 1 & 1 \\[3 pt]

Unidentified & --   & 8 \\[3 pt]

\hline

\end{tabular}

\end{center}

\caption{Census of the 47 First AGILE High-Confidence Gamma-Ray Sources.} 
\label{tab:assoc}

\end{table*}

In Table \ref{tab:assoc}, for  ``confirmed" counterparts is meant  
$\gamma$-ray sources for which there
are peer reviewed publications demonstrating high-confidence
association with refined analysis methods. 
Associations for uncertain sources have been
selected using cross-correlations with various updated public catalogs
of $\gamma$, hard X-ray and radio sources
either of specific mission or of specific source classes, such as:
\begin{itemize}
\item the Third EGRET Catalog (3EG) \citep{1999ApJS..123...79H}
and the EGRET Revised Catalog of Gamma-Ray Sources \citep{2008A&A...489..849C};
\item the INTEGRAL Reference Catalog (INTREFCAT) \citep{2003A&A...411L..59E};
\item a selection from the Australian Telescope National Facility (ATNF) Pulsar Catalog
\citep{ATNF};
\item H.E.S.S source catalog (available on-line);
\item the SNR Catalog \citep{Gr91,Gr09};
\item the Blazar Roma-BZCAT \citep{BZCAT}. 
\end{itemize}




Summing up, the first AGILE {\nnew catalog} 
includes:
21 confirmed or candidate pulsars, 13 Blazars
(7 FSRQ, 4 BL Lacs, 2 unknown type), 2 HMXRBs, 2 SNRs, 1 colliding-wind binary system (CWB)
and 8 unidentified sources.
\par


In Table 3  we report the values of the following relevant
source parameters:
\begin{itemize}
\item{} AGILE source name,
\item{} Source position both in Celestial and Galactic sky coordinates: RA, DEC (J2000) and LII, BII
\item{} Position Error (95\%), defined as the 2-D
error circle radius at 95\% confidence level, statistical error only\footnote{
The AGILE Team recommends to add linearly to this value a systematic error of $\pm 0.1^{\circ}$.}.
\item{} The $\sqrt{TS}$ values of the  significance of the detection as determined from the refined
ML analysis.
\item{}  The mean value of the F4 exposure map in units of 10$^{8}$ cm$^{2}$ s,
relative to the sky area (Ring) used for each source analysis.
\item{}  The source flux above 100 MeV and its $1 \sigma$ statistical error\footnote{
The AGILE Team recommends to add to the flux statistical error a systematic error of $10\%$.}
in units of $10^{-8}$ ph cm$^{-2}$ s$^{-1}$.
This is the average source flux value over the entire time period.
\item{}  Source classification.
\item{}  Counterpart name for confirmed sources.
\item{}  Possible Counterparts in the AGILE error radius and Other Names,
both for ``confirmed'' and ``uncertain'' counterparts.
 \end{itemize}


\subsection{Notes on {\cp individual} sources}
\par
As described in Sect.\ref{sec:method},
pointlike $\gamma$-ray sources {\cp parameters} reported in this paper are determined by a 
likelihood analysis of the $10^{\circ}$ field surrounding the candidate sources. 
The analysis depends on the local Galactic diffuse emission, the $\gamma$-ray photon 
statistics, the instrument PSF, the response matrix as a function of energy and off-axis angle, 
and on the background filtering. 
{\cp Particular care is required to carry out the analysis}
in regions of the Galactic plane that are characterized by a relatively high and structured 
flux of the diffuse Galactic emission, {\cp as well as in regions harboring
nearby $\gamma$-ray sources leading to possible source confusion.}
For such regions 
we insert the label (C), for ``Confused", in the Confirmed Counterpart column
of Table 3.  
Note that these are significant AGILE detections which however
have flux and location parameters that may be affected
(within the statistical+systematic errors)
by other nearby sources.
In the following we briefly comment on some specific AGILE detections.
~
\par \noindent
~
\par \noindent
{\bf 1AGL J0006+7311}: This AGILE source, positionally coincident with
the EGRET $\gamma$-ray source 3EG J0010+7309 located in
supernova remnant CTA 1, is associated with the first radio quiet pulsar
recently discovered through its $\gamma$-ray pulsations by the Fermi Gamma-Ray Space Telescope
~\citep{Abdo}.
This new class of young pulsar sources may be possibly
associated with most unidentified Galactic $\gamma$-ray sources in  star-forming regions 
and SNRs.
Search for pulsations in $\gamma$-ray AGILE data is currently under way.
{\cp At the border of the AGILE error box there is also the Blazar source BZQ J0019+7327.}
\\
\par \noindent
 {\bf1AGL J0535+2205 and 1AGL J0634+1748 (Crab and Geminga)}: These two well known strong $\gamma$-ray
 pulsars, together with the Vela pulsar,
 were used for in-flight AGILE calibrations. We report in table the flux values obtained
during calibration sub-periods. These values are in agreement with pulsed flux values reported in
\citep{pellizzoni08}. 
We note however that we observed higher flux values,
over $1 \sigma$  from the reported mean flux, for both sources {\cp when merging}
all the data, including shorter (1 day) integration periods during 2007. This
{\cp point} is under investigation.
\\
\par \noindent
{\bf 1AGL J0617+2236}: This AGILE detection provides an improved positioning
compared to the 3EG J0617+2238 error box.
{\cp This source is positionally
coincident with the SNR IC443 \cite{TavaniIC443}. The AGILE error box  contains also}
the PSR J0614+2229.
\\
\par \noindent
{\bf  1AGL J0657+4554 and 1AGL J0714+3340}: These two high latitude ($|b|>10$ deg) AGILE
sources,  {\cp associated with Blazar of unknown type in the BZCAT,
have no EGRET counterparts probably owing to flux variability}.
\\
\par \noindent
{\bf 1AGL J0835-4509 (Vela pulsar)}:  {\cp As the most luminous 
steady source in the $\gamma$-ray sky,
Vela has been extensively used
for in-flight AGILE-GRID calibrations}. Note that with the F4  filter version 
and the strict criteria 
used to build this first catalog, the resulting effective exposure is 
quite low (only about $0.81
\times 10^8$ cm$^2$ sec on source over the entire period).
\\
\par \noindent
{\bf 1AGL J1022-5822}: This  {\cp source lies}
in the complex Carina region, multiple source contributions
are possible.
\\
\par \noindent
{\bf 1AGL J1043-5931}:   {\new This source (not detected by EGRET)
is close to 1AGL J1043-5749 in the Carina region. Our refined
analysis leads to the association of this $\gamma$-ray source with
the colliding wind binary  Eta Carinae \cite{Tavanietacar}.
\\
\par \noindent
 {\bf 1AGL J1104+3754 and 1AGL J1222+2851}: The effective exposure on these sources is low, just about 2 effective days, but  it includes a ToO period on the source W Comae.
 \\
 \par \noindent
 {\bf 1AGL J1412-6149 and 1AGL J1419-6055}: This  {\cp source lies}
in the complex Crux region, multiple source contributions are possible.
 \\
\par \noindent
 {\bf 1AGL J1511-0908}: The total effective exposure on this source is very low,
just about 2 effective days,
but it includes a ToO period on the associated source PKS 1510-089.
\\
\par \noindent
 {\bf 1AGL J1736-3235,  1AGL J1746-3017}:
 {\new These sources are in the complex region around $10^{\circ}$ 
from the Galactic Center, multiple source contributions are possible.
We emphasize the relatively small exposure of the Galactic Center
region achieved until June 30, 2008 that does not allow a 
deeper analysis of the complex $\gamma$-ray emission from the center 
of our Galaxy.}
\\
\par \noindent
{\bf 1AGL J1801-2317}:   {\new This source is spatially coincident
with the TeV source HESS~J1801-233. Remarkably, they both
appear to be associated with the Northeastern section of the SNR
W28 shell \cite{giulianiw28}.}


\section{Conclusions}
\label{sec:concl}

The AGILE Cycle-1 Pointing Plan
 {\cp covered the whole sky focusing mainly toward
the Galactic plane.} 
The AGILE first catalog includes only high-significance sources characterized by a prominent mean 
$\gamma$-ray flux above 100 MeV when integrated over the total exposure period 2007 July - 2008 June.
With  {\cp our one-year long integration}
only sources with ``steady" flux values above $\sim 20 \times 10^{-8}$ ph cm$^{-2}$ s$^{-1}$
are detected over $ 4 \sigma$. Source detections
during flaring state and determination of peak fluxes are not included in this catalog.
This should be taken into account when comparing with
the results of the third EGRET catalog which includes detections over  $ 4 \sigma$
in each of the EGRET viewing periods during its effective 6-year lifetime.
An analysis of $\gamma$-ray detection by the AGILE-GRID over short timescales 
(several weeks, 1-week, days) is beyond the scope of this catalog and it will 
be published elsewehere.  The AGILE-GRID spatial resolution reached with long 
exposures is substantially better than that of EGRET, and the total exposure 
accumulated by AGILE in several sky regions,
particularly near the Galactic plane, is comparable with that obtained by
EGRET in 6 years effective time. 
It is then interesting to compare the relatively high-flux sources 
detected by AGILE with the equivalent  sources of the third EGRET catalog. 
Many bright $\gamma$-ray sources detected by EGRET are confirmed by {\cp AGILE which
 provides comparable or better positioning.}
AGILE in the first catalog detected five sources that were not present
in the 3EG catalog: 3 Blazars and 2 candidate pulsars.
As expected due to statistical detection effects and source variability, some of the prominent (with flux range 40-100 $\times 10^{-8}$ ph cm$^{-2}$ s$^{-1}$)
3EG  sources in the Galactic plane are not detected by AGILE with mean flux values at a significance level that is sufficient to be included in this first catalog.

It is also important to note that the AGILE-GRID exposure in the
selected period has been accumulated mostly in the Carina-Crux and in the Cygnus regions, with relatively low exposure at the Galactic center. This explains the relatively small number of sources in the Galactic center region.

Finally, taking into account that the AGILE first catalog is not a complete
flux-limited sample 
and is affected by selection effects  due to the assumed fixed value ($-2.1$) of
the unknown source spectral {\cp indices},
we observe that with a limiting flux of about 2 $\times 10^{-7}$ ph cm$^{-2}$ s$^{-1}$,
the number and rate of  $\gamma$-ray Blazars  observed by AGILE
(13 Blazars: 7 FSRQs and 5 BL Lacs) is roughly consistent
with expectations from the EGRET log $N$-log $S$ \citep{Ozel,Mucke}.
A detailed study over a complete AGILE AGN sample will be performed in the future.

A variability study of the sources of this first catalog over different timescales
will appear in Verrecchia et al. 2009.

\begin{acknowledgements}
The AGILE Mission is funded by the Italian Space Agency (ASI) with
scientific and programmatic participation by the Italian Institute
of Astrophysics (INAF) and the Italian Institute of Nuclear
Physics (INFN).  We acknowledge funds from ASI grant I/089/06/2.

\end{acknowledgements}

\bibliographystyle{aa} 
\bibliography{AGILE_first_catalog_revised}





\clearpage

\begin{figure}
\twocolumn
\centering
\vskip -6truecm
\hbox{
\includegraphics[width=1.2\textwidth]{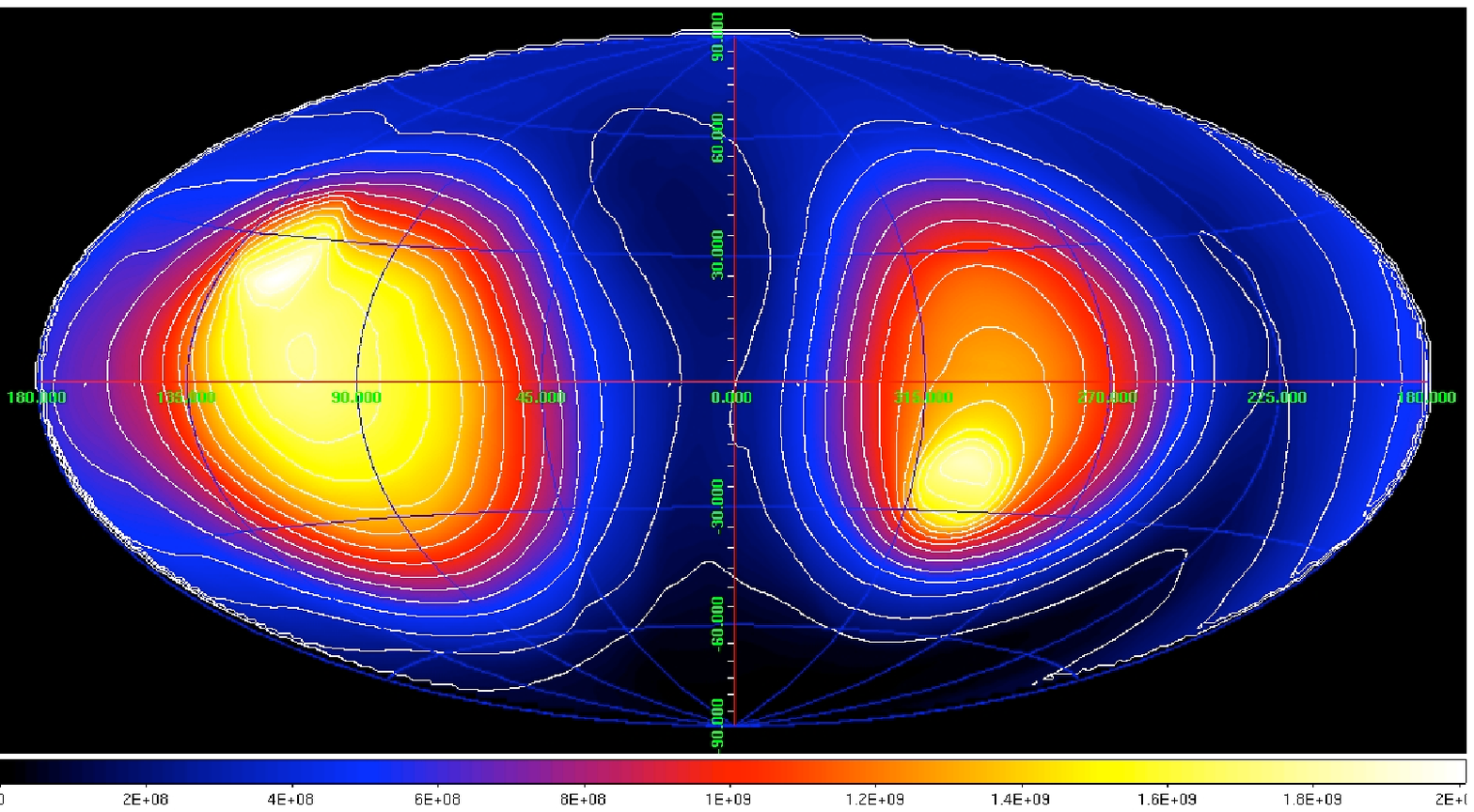}
}
\onecolumn
\vskip -7truecm
\caption{{\new Total AGILE-GRID exposure sky map in Aitoff projection and
Galactic coordinates, for energies $>100$ \Mev in
units of cm$^2$ s,  accumulated during the period July 9, 2007 - June 30, 2008
(with the F4 event filter).
The regions of deeper exposures (whiter in the color scale) are a consequence of the AGILE
specific pointings at the Galactic plane, combined with the effect of Earth occultation.}
}
\label{fig:expo}
\end{figure}


\begin{figure}
\twocolumn
\centering
\hbox{
\includegraphics[width=\textwidth]{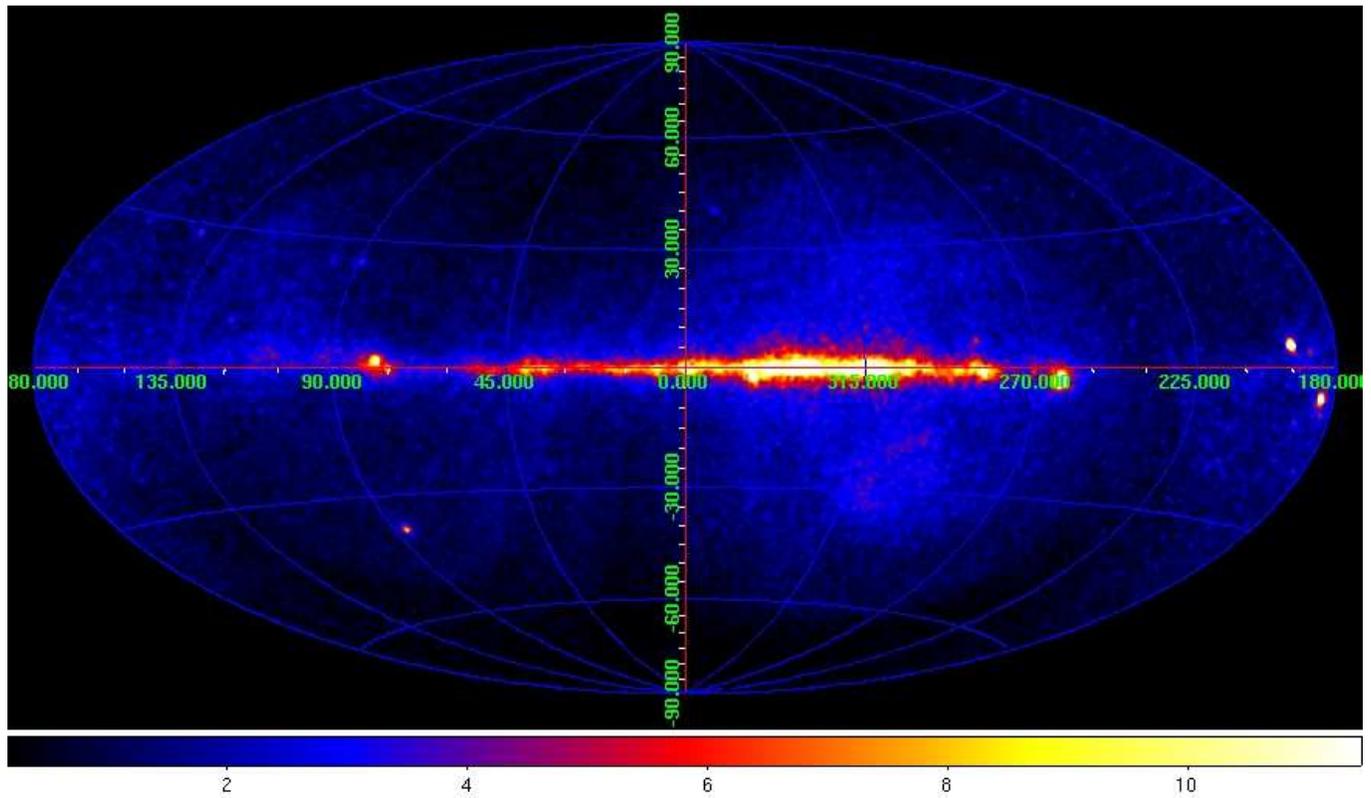}
}
\onecolumn
\caption{{\new Total AGILE-GRID count map in Aitoff projection and
Galactic coordinates, for energies $>100$ \Mev in
units of ph cm$^{-2}$ s$^{-1}$  sr$^{-1}$
accumulated during the period July 9, 2007 - June 30, 2008
(with the F4 event filter).
The effect of the non-uniform exposure is particularly evident
for pointings centered near the Carina-Crux region.}
}
\label{fig:counts}
\end{figure}

\clearpage

\begin{figure}
\centering
\hbox{
\includegraphics[angle=90,width=\textwidth]{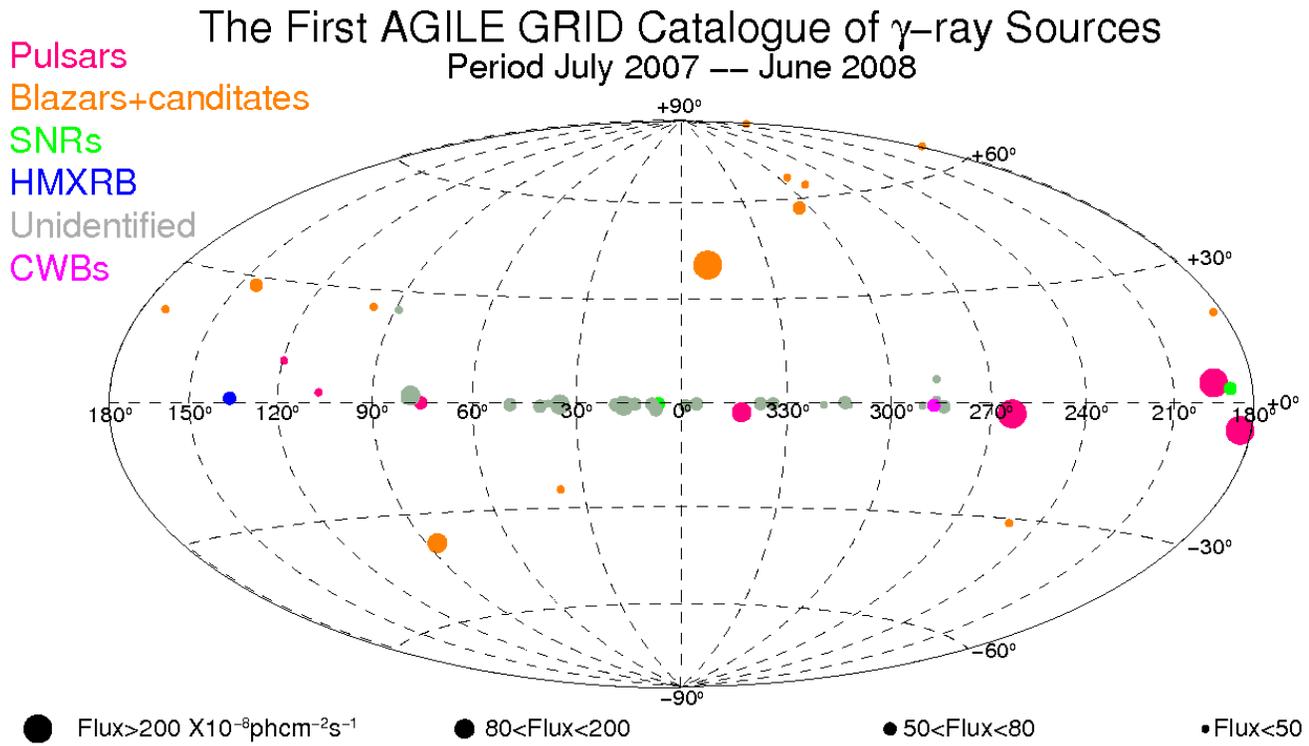}
} \onecolumn \caption{The First AGILE-GRID Catalog of
high-confidence sources, plotted in in Aitoff projection and galactic sky coordinates.
Symbol sizes are proportional to source flux values and symbol colors indicate different
source classes.}
\label{fig:aitoff}
\end{figure}



\input{pointinglist_63_v10.tex}

\clearpage

\input{first_cat_table_final_v8.tex}

\clearpage



\end{document}

%% file: pointinglist_63_v10.tex
\begin{table*}[tbh]

\hskip 2truecm
\scriptsize

\begin{center}


\begin{tabular}{|lccccc|}

\hline
 & & & & & \\
\multicolumn{1}{|l}{Region Name} &
\multicolumn{1}{c}{OB Number} &
\multicolumn{1}{c}{Starting RA, Dec  J2000 (deg)} &
\multicolumn{1}{c}{Starting LII, BII (deg)} &
\multicolumn{1}{c}{Start Observation (UTC)} &
\multicolumn{1}{c|}{End Observation(UTC)} \\
 & & & & & \\
\hline
3C279 Region  &  OB900      &  195.596  ,  -6.649   &    307.8118 ,  56.1183 &	 2007-07-09 12:00  &  2007-07-13 12:00     \\
VELA Region   &  OB1000     &  157.979  ,  -60.214  &    286.4188 ,  -1.8951 &	 2007-07-13 12:00  &  2007-07-24 12:00	   \\
ToO 3C 454.3  &  OB1100     &  17.829   ,  36.694   &    127.3645 , -26.0059 &	 2007-07-24 12:00  &  2007-07-30 12:00	   \\
ToO 3C 454.3  &  OB1150     &  17.829   ,  36.694   &    127.3645 , -26.0059 &	 2007-07-24 12:00  &  2007-07-30 12:00	   \\
VELA Region   &  OB1200     &  150.836  , -70.19   &    289.5293 , -11.8265 &	 2007-07-30 12:00  &  2007-08-01 12:00	   \\
SA Crab -45   &  OB1300     &  37.097   , 12.712   &    156.5885 , -43.7329 &	 2007-08-01 12:00  &  2007-08-02 12:00	   \\
VELA Region   &  OB1400     &  176.006  , -66.063  &    296.1593 ,  -4.0824 &	 2007-08-02 12:00  &  2007-08-12 12:00	   \\
SA Crab -35   &  OB1500     &  47.41    , 16.075   &    164.8343 , -35.3162 &	 2007-08-12 12:00  &  2007-08-13 12:00	   \\
VELA Region   &  OB1600     &  195.551  , -66.564  &    304.0044 ,  -3.7154 &	 2007-08-13 12:00  &  2007-08-22 12:00	   \\
SA Crab -25   &  OB1700     &  57.139   , 18.566   &    171.0790 , -27.3115 &	 2007-08-22 12:00  &  2007-08-23 12:00	   \\
VELA Region   &  OB1800     &  216.979  , -64.437  &    313.1071 ,  -3.4890 &	 2007-08-23 12:00  &  2007-08-27 12:00	   \\
Galactic Plane  &  OB1900   &  236.570  , -41.874  &    334.4369 ,  10.0581 &	 2007-08-27 12:00  &  2007-09-01 12:00	   \\
SA Crab (15,15) &  OB2000   &  69.483  , 5.592    &    190.8962 , -26.2858 &	 2007-09-01 12:00  &  2007-09-02 12:00	   \\
SA Crab (0,15)  &  OB2100   &  68.205   , 20.566   &    177.1349 , -18.2781 &	 2007-09-02 12:00  &  2007-09-03 12:00	   \\
SA Crab (-15,15)&  OB2200   &  66.651   , 35.559   &    164.6334 ,  -9.3529 &	 2007-09-03 12:00  &  2007-09-04 12:00	   \\
Field 8        &  OB2300    &  51.408   , 71.022   &    134.8816 ,  11.8210 &	 2007-09-04 12:00  &  2007-09-12 12:00	   \\
SA Crab (0,5)  &  OB2400    &  78.535   , 21.730   &    182.1630 ,  -9.8874 &	 2007-09-12 12:00  &  2007-09-13 12:00	   \\
Field 8        &  OB2500    &  74.882   , 58.334   &    150.9906 ,   9.7255 &	 2007-09-13 12:00  &  2007-09-15 12:00	   \\
SA Crab (45,0)  &  OB2600   &  84.212   , -23.014  &    226.7035 , -26.1161 &	 2007-09-15 12:00  &  2007-09-16 12:00	   \\
SA Crab (5,0)  &  OB2700    &  82.987   , 16.983   &    188.5217 ,  -8.9833 &	 2007-09-16 12:00  &  2007-09-17 12:00	   \\
SA Crab (0,0)  &  OB2800    &  83.774   , 22.026   &    184.6179 ,  -5.6675 &	 2007-09-17 12:00  &  2007-09-18 12:00	   \\
SA Crab (-5,0)   &  OB2900  &  84.62    , 27.048   &    180.7737 ,  -2.3343 &	 2007-09-18 12:00  &  2007-09-19 12:00	   \\
SA Crab (-15,0)  &  OB3000  &  85.347   , 37.089   &    172.5873 ,   3.5179 &	 2007-09-19 12:00  &  2007-09-20 12:00	   \\
SA Crab (-25,0)  &  OB3100  &  86.174   , 47.118   &    164.2603 ,   9.2213 &	 2007-09-20 12:00  &  2007-09-21 12:00	   \\
SA Crab (-35,0)  &  OB3200  &  87.140   , 57.126   &    155.6110 ,  14.6016 &	 2007-09-21 12:00  &  2007-09-22 12:00	   \\
SA Crab (-45,0)  &  OB3300  &  88.348   , 67.136   &    146.4473 ,  19.4825 &	 2007-09-22 12:00  &  2007-09-23 12:00	   \\
SA Crab (0,-5)   &  OB3400  &  90.097   , 22.143   &    187.5419 ,  -0.5862 &	 2007-09-23 12:00  &  2007-09-24 12:00	   \\
SA Crab (15,0)   &  OB3500  &  91.034   , 7.141    &    201.1056 ,  -7.1395 &	 2007-09-24 12:00  &  2007-09-25 12:00	   \\
SA Crab (25,0)   &  OB3600  &  91.838   , -2.882   &    210.4602 , -11.1195 &	 2007-09-25 12:00  &  2007-09-26 12:00	   \\
SA Crab (35,0)   &  OB3700  &  92.502   , -12.926  &    220.0176 , -14.9489 &	 2007-09-26 12:00  &  2007-09-27 12:00	   \\
Crab Nebula      &  OB3800  &  94.323   , 22.050   &    189.5211 ,   2.7938 &	 2007-09-27 12:00  &  2007-10-01 12:00	   \\
SA Crab (0,-15)  &  OB3900  &  98.552   , 21.875   &    191.4932 ,   6.1922 &	 2007-10-01 12:00  &  2007-10-02 12:00	   \\
SA Crab (-15,-15)&  OB4000  &  100.839  , 36.784   &    178.6417 ,  14.3544 &	 2007-10-02 12:00  &  2007-10-03 12:00	   \\
SA Crab (15,-15) &  OB4100  &  99.566   , 6.788    &    205.3927 ,   0.1791 &	 2007-10-03 12:00  &  2007-10-04 12:00	   \\
Crab Field       &  OB4200  &  101.724  , 21.699   &    192.9681 ,   8.7550 &	 2007-10-04 12:00  &  2007-10-12 12:00	   \\
SA Crab (0,-25)  &  OB4300  &  110.131  , 20.718   &    197.2281 ,  15.4667 &	 2007-10-12 12:00  &  2007-10-13 12:00	   \\
Gal. Center      &  OB4400  &  290.920  , -18.896  &     19.2683 , -15.4110 &	 2007-10-13 12:00  &  2007-10-22 12:00	   \\
SA Crab (0,-35)  &  OB4500  &  120.494  , 18.879   &    203.0392 ,  23.7444 &	 2007-10-22 12:00  &  2007-10-23 12:00	   \\
Gal. Center Reg. &  OB4600  &  301.173  , -17.107  &     25.0972 , -23.6663 &	 2007-10-23 12:00  &  2007-10-24 08:00	   \\
ToO 0716+714     &  OB4610  &  148.939  , 67.888   &    143.3642 ,  41.5875 &	 2007-10-24 08:00  &  2007-10-29 12:00	   \\
ToO Extended  &  OB4630  &  157.461  , 66.942   &    141.5537 ,  44.7248 &	 2007-10-29 12:00  &  2007-11-01 12:00	   \\
SA Crab (0,-45)  &  OB4700  &  130.614  , 16.339   &    209.7914 ,  31.7351 &	 2007-11-01 12:00  &  2007-11-02 12:00	   \\
Cygnus Region    &  OB4800  &  296.880  , 34.501   &     69.5937 ,   4.6227 &	 2007-11-02 12:00  &  2007-12-01 12:00	   \\
Cygnus Field 1   &  OB4900  &  304.432  , 53.552   &     88.8156 ,   9.9272 &	 2007-12-01 12:00  &  2007-12-05 09:00	   \\
Cygnus Repointing&  OB4910  &  322.496  , 38.244   &     85.1187 ,  -9.4171 &	 2007-12-05 09:00  &  2007-12-16 12:00     \\
Cygnus Repointing&  OB4920  &  322.496  , 38.244   &     85.1187 ,  -9.4171 &	 2007-12-05 09:00  &  2007-12-16 12:00     \\
Virgo  Field     &  OB5010  &  173.433  , -0.437   &    265.6464 ,  56.7005 &	 2007-12-16 12:00  &  2008-01-08 12:00	   \\
Vela  Field      &  OB5100  &  147.060  , -62.517  &    283.4703 ,  -6.7881 &	 2008-01-08 12:00  &  2008-02-01 12:00	   \\
South Gal Pole   &  OB5200  &   58.347  , -37.795  &    240.3889 , -50.5780 &	 2008-02-01 12:00  &  2008-02-09 09:00	   \\
ToO MKN 421      &   OB5210 &  250.974  ,  50.293  &     77.3096 ,  40.6278 &	 2008-02-09 09:00  &  2008-02-12 12:00	   \\
South Gal Pole Repointing&OB5220&65.660 , -35.714  &    237.5007 , -44.6737 &	 2008-02-12 12:00  &  2008-02-14 12:00     \\
Musca Field      &  OB5300  &  191.934  , -71.893  &    302.6408 ,  -9.0241 &	 2008-02-14 12:00  &  2008-03-01 12:00	   \\
Gal. Center 1    &  OB5400  &  243.596  , -50.979  &    332.1063 ,   0.0207 &	 2008-03-01 12:00  &  2008-03-16 12:00	   \\
Gal. Center 2    &  OB5450  &  265.781  , -28.626  &    359.9782 ,   0.6280 &	 2008-03-16 12:00  &  2008-03-30 12:00	   \\
Anti-Center 1    &  OB5500  &  100.944  ,  21.711  &    192.6369 ,   8.1084 &	 2008-03-30 12:00  &  2008-04-05 12:00	   \\
SA Crab (8,24)   &  OB5510  &  108.283  ,  28.625  &    188.9607 ,  16.9953 &	 2008-04-05 12:00  &  2008-04-07 12:00	   \\
SA Crab (15,26)  &  OB5520  &  111.762  ,  35.688  &    183.0072 ,  22.2023 &	 2008-04-07 12:00  &  2008-04-08 12:00	   \\
Anti-Center 2    &  OB5530  &  110.404  ,  20.758  &    197.2962 ,  15.7167 &	 2008-04-08 12:00  &  2008-04-10 12:00	   \\
Vulpecula Field  &  OB5600  &  286.259  ,  20.819  &     53.0394 ,   6.4733 &	 2008-04-10 12:00  &  2008-04-30 12:00	   \\
North Gal Pole   &  OB5700  &  250.075  ,  72.497  &    104.8522 ,  35.4379 &	 2008-04-30 12:00  &  2008-05-10 12:00	   \\
Cygnus Field 2   &  OB5800  &  304.286  ,  35.974  &     74.0497 ,   0.2720 &	 2008-05-10 12:00  &  2008-06-09 18:00	   \\
ToO WComae ON+231&  OB5810  &  182.285  ,  29.614  &    195.5016 ,  80.3738 &	 2008-06-09 18:00  &  2008-06-15 12:00     \\
Cygnus Repointing&  OB5820  &  323.248  ,  50.079  &     93.6645 ,  -1.1664 &	 2008-06-15 12:00  &  2008-06-30 12:00     \\
\hline

\end{tabular}

\end{center}

\normalsize
\caption{AGILE pointings in the period 9 July 2007 - 30 June 2008, corresponding to 
63 Observation Blocks (OB) considered in our analysis.
Acronyms used in table are: ToO = Target of Opportunity pointing, SA = SuperAGILE special pointing.}

\label{table:pointings}
\end{table*}

%% file: first_cat_table_final_v8.tex
\newcommand{\bold}{}


\begin{landscape}

\markright{}
\authorrunning{}
\titlerunning{}

\begin{table*}[tbh]


\scriptsize


\begin{tabular}{|lccccccccccc|}

\hline
 & & & &  & & & & & & & \\
\multicolumn{1}{|l}{AGILE name} & 
\multicolumn{1}{c}{RA (J2000.0)} & 
\multicolumn{1}{c}{Dec (J2000.0)} & 
\multicolumn{1}{c}{LII} &
\multicolumn{1}{c}{BII} &
\multicolumn{1}{c}{$^{a}$Pos. Error (95\%)} & 
\multicolumn{1}{c}{sqrt(TS)} & 
\multicolumn{1}{c}{$^{b}$Mean Ring Exp} &
\multicolumn{1}{c}{$^{c}$Mean Flux \& Error} & 
\multicolumn{1}{c}{Classification} &
\multicolumn{1}{c}{Confirmed Counterp.} &
\multicolumn{1}{c|}{Possible Counterp. }      \\[3 pt]
\multicolumn{1}{|l}{        } & 
\multicolumn{1}{c}{  (hh mm ss) } & 
\multicolumn{1}{c}{  (dd mm ss) } & 
\multicolumn{1}{c}{ (deg) } &
\multicolumn{1}{c}{ (deg) } &
\multicolumn{1}{c}{ (deg)} & 
\multicolumn{1}{c}{   } & 
\multicolumn{1}{c}{  ($\times 10^{8}$ cm$^{2}$ s) } &
\multicolumn{1}{c}{  ($\times 10^{-8}$ ph cm$^{-2}$ s$^{-1}$) } & 
\multicolumn{1}{c}{           } &
\multicolumn{1}{c}{           } &
\multicolumn{1}{c|}{\& Other Names}      \\[3 pt]
\hline
 & & & &  & & & & & & & \\[3 pt]
1AGL J0006+7311  &   00 06 34.2  &   +73 11 06.6 &   119.65   &   10.6     &    0.63    &  5.1  &  3.01  & 23 $\pm$ 5  &   GammaPulsar*   &   CTA1     &   3EGJ0010+7309   \\[7 pt] 
1AGL J0242+6111  &   02 42 13.6  &   +61 11 06.7 &   135.88   &   1.13     &    0.64    &  5.3  &  1.17  & 54 $\pm$ 12  &   HMXRB          &   LSI+61303     &     3EGJ0241+6103 \\[7 pt] 
1AGL J0535+2205  &   05 35 05.9  &   +22 05 41.7 &   184.56   &   -5.63    &    0.09    &  47.2 &  2.79  & 220 $\pm$ 15  &   Pulsar         &   Crab          &     3EGJ0534+2200 \\[7 pt]  
1AGL J0538-4424  &   05 38 29.6  &   -44 24 17.8 &   250.44   &   -31.2    &    0.5     &  5.9  &   0.81  & 43 $\pm$ 10  &   Blazar-BLLac   &   PKS0537-441   &     3EGJ0540-4402 \\  
 & & & &  & & & & & & & BZBJ0538-4405 \\[7 pt]
1AGL J0617+2236  &   06 17 21.7  &   +22 36 14.2 &   189.04   &   3.07     &    0.27    &  9.9  &  2.79  & 69 $\pm$ 9  &   {\bold SNR}   &   {\bold IC443 }        &     3EGJ0617+2238 \\[7 pt]
1AGL J0634+1748  &   06 34 15.8  &   +17 48 27.7 &   195.14   &   4.36     &    0.05    &  63   &  2.79  & 320 $\pm$ 10  &   Pulsar         &   Geminga       &     3EGJ0633+1751 \\[7 pt]  
1AGL J0657+4554  &   06 57 29.2  &   +45 54 14.5 &   170.73   &   20.11    &    0.55    &  5.8  &  1.98  & 31 $\pm$ 6  &   Blazar        &   ---           &  BZUJ0654+4514 \\  
& & & & &  & & & & & & S40650+45 \\[7 pt]
1AGL J0714+3340  &   07 14 29.4  &   +33 40 37.3 &   184.12   &   19.1     &    0.85    &  4.2  &  2.57  & 18 $\pm$ 5  &   Blazar        &   ---           &     BZUJ0719+3307  \\
 & & & &  & & & & & & & GB20716+332  \\[7 pt] 
1AGL J0722+7125  &   07 22 22.9  &   +71 25 31.1 &   143.89   &   28.06    &    0.37    &  10.9 &  1.39  & 68 $\pm$ 9  &   Blazar-BLLac   &   S50716+714    &     3EGJ0721+7120 \\
 & & & &  & & & & & & & BZBJ0721+7120 \\[7 pt]
1AGL J0835-4509  &   08 35 13.3  &   -45 09 09.0 &   263.52   &   -2.79    &    0.09    &  41.7 &   0.81  & 780 $\pm$ 32  &   Pulsar         &   VelaPSR       &     3EGJ0834-4511 \\[7 pt]  
1AGL J1022-5822  &   10 22 08.8  &   -58 22 17.0 &   284.39   &   -0.98    &    0.36    &  10.1 &  4.85  & 59 $\pm$ 7  &   Unclassified   &    (C)           &     3EGJ1013-5915 \\
 & & & &  & & & & & & & PSRJ1016-5857 \\[7 pt]
1AGL J104{\bold 4}-593{\bold 6}  &  10 43 {\bold 57.6}    &  -59 3{\bold 6 41.3} &   287.{\bold 4}4   &   -0.{\bold 63}    &   0.{\bold 58}	  &  5.{\bold 3} &  4.85  & 2{\bold 7} $\pm$ 6  &   CWB   &    {\bold ---  }          &   {\bold EtaCar } \\[7 pt]   

{\bold 1AGL J1043-5749  }& {\bold 10 43 56.0}    &  {\bold -57 49 51.0} &   {\bold 286.60}   &   {\bold 0.94}    &  {\bold 0.68}	  & {\bold 4.5} & {\bold 4.85}  &{\bold 22 $\pm$ 5}  &  {\bold Unclassified }  &  {\bold (C)  }         &  {\bold 3EGJ1048-5840}   \\[7 pt] 

1AGL J1058-5239  &  10 58 31.1    &  -52 39 47.5 &   286.15   &   6.49     &   0.30     &  8.7 &  4.85  & 29 $\pm$ 4  &   Unclassified   &  ---	    &     3EG J1058-5234 \\
& & & & &  & & & & & & PSRJ1057-5226 \\[7 pt]
1AGL J1104+3754  &   11 04 38.5  &   +37 54 33.6 &   180.48   &   65.16    &    0.66    &  4.7  &   0.51  & 42 $\pm$ 13  &   Blazar-BLLac   &   Mkn421        &     3EGJ1104+3809 \\ 
& & & & &  & & & & & & BZBJ1104+3812 \\[7 pt]
1AGL J1108-6103  &   11 08 43.6  &   -61 03 54.3 &   290.83   &   -0.63    &    0.57    &  6.1  &  4.85  & 30 $\pm$ 6  &   Unclassified   &   ---           &     3EGJ1102-6103 \\  
& & & & &  & & & & & & PSRJ1119-6127 \\[7 pt]
1AGL J1222+2851  &   12 22 39.7  &   +28 51 02.3 &   196.09   &   83.42    &    0.74    &  4.7  &   0.50  & 38 $\pm$ 11  &   Blazar-BLLac   &   WComae        &     3EGJ1222+2841 \\  
& & & & &  & & & & & &  BZBJ1221+2813 \\
& & & & &  & & & & & & ON +231 \\[7 pt]
1AGL J1228+0142  &   12 28 59.5  &   +01 42 41.3 &   290.04   &   64.02    &    0.71    &  4.7  &  1.98  & 24 $\pm$ 6  &   Blazar-FSRQ    &   3C273         &     3EGJ1229+0210 \\ 
& & & & &  & & & & & &  BZQJ1229+0203 \\[7 pt]
1AGL J1238+0406  &   12 38 31.0  &   +04 06 14.2 &   294.74   &   66.77    &    1.23    &  4.7  &  1.98  & 25 $\pm$ 6  &   Blazar-FSRQ    &   ---           &     3EGJ1236+0457 \\ 
& & & & &  & & & & & &  BZQJ1239+0443 \\[7 pt]
1AGL J1256-0549  &   12 56 33.1  &   -05 49 42.6 &   305.27   &   57.02    &    0.32    &  10.2 &  1.98  & 65 $\pm$ 9  &   Blazar-FSRQ    &   3C279         &     3EGJ1255-0549 \\
& & & & &  & & & & & & BZQJ1256-0547  \\[7 pt]
1AGL J1412-6149 &   14 12 06.1   &  -61 49 32.5   &   312.3    &   -0.43   &   0.44    &   6.3  &  5.44  & 43 $\pm$ 7  &  Unclassified &    (C)   &  3EGJ1410-6147 \\
& & & & &  & & & & & &  PSRJ1410-6132 \\
& & & & &  & & & & & & G312.4-0.4 \\[7 pt]
\hline

\end{tabular}


\normalsize
\caption{First AGILE High Confidence Gamma-Ray Sources List.}

\end{table*}

\begin{table*}[tbh]


\scriptsize


\begin{tabular}{|lccccccccccc|}

\hline
 & & & &  & & & & & & & \\
\multicolumn{1}{|l}{AGILE name} & 
\multicolumn{1}{c}{RA (J2000.0)} & 
\multicolumn{1}{c}{Dec (J2000.0)} & 
\multicolumn{1}{c}{LII} &
\multicolumn{1}{c}{BII)} &
\multicolumn{1}{c}{$^{a}$Pos. Error (95\%)} & 
\multicolumn{1}{c}{sqrt(TS)} & 
\multicolumn{1}{c}{$^{b}$Mean Ring Exp} &
\multicolumn{1}{c}{$^{c}$Mean Flux \& Error} & 
\multicolumn{1}{c}{Classification} &
\multicolumn{1}{c}{Confirmed Counterp.} &
\multicolumn{1}{c|}{Possible Counterp. }      \\[3 pt]
\multicolumn{1}{|l}{        } & 
\multicolumn{1}{c}{  (hh mm ss) } & 
\multicolumn{1}{c}{  (dd mm ss) } & 
\multicolumn{1}{c}{ (deg) } &
\multicolumn{1}{c}{ (deg) } &
\multicolumn{1}{c}{ (deg)} & 
\multicolumn{1}{c}{   } & 
\multicolumn{1}{c}{  ($\times 10^{8}$ cm$^{2}$ s) } &
\multicolumn{1}{c}{  ($\times 10^{-8}$ ph cm$^{-2}$ s$^{-1}$) } & 
\multicolumn{1}{c}{           } &
\multicolumn{1}{c}{           } &
\multicolumn{1}{c|}{\& Other Names}      \\[3 pt]
\hline
 & & & &  & & & & & & & \\[3 pt]
1AGL J1419-6055 &   14 19 51.2   &  -60 55 11.2   &   313.47   &   0.13    &   0.31    &   7.5  &  5.44  & 52 $\pm$ 7  &  Unclassified &    (C)   &  3EGJ1420-6038 \\
& & & & &  & & & & & &  PSRJ1420-6048 \\[7 pt]
1AGL J1506-5859 &   15 06 01.5   &  -58 59 13.5   &   319.52   & -0.52     &   0.48    & 6.9   &  5.44  & 41 $\pm$ 7  &  Unclassified &  --- &  PSRJ1509-5850 \\[7 pt]
1AGL J1511-0908  &   15 11 38.5  &   -09 08 12.8 &   350.97   &   40.31    &    0.33   &  11.2 &   0.39  & 220 $\pm$ 32  &   Blazar-FSRQ    &   PKS1510-089  &     3EGJ1512-0849 \\
& & & & &  & & & & & & BZQJ1512-0905 \\[7 pt]
1AGL J1624-4946  &   16 24 26.9  &  -49 46 51.9  &  334.09    &  -0.25    &    0.58  &  5.7  &  2.18  & 67 $\pm$ 13  &  Unclassified &  --- &  PSRJ1623-4949  \\[7 pt] 
1AGL J1639-4702  &   16 39 05.5  &  -47 02 28.2  &  337.75    &  -0.15    &    0.53  &  6.4  &  2.18  & 76 $\pm$ 13  &  Unclassified &  --- &  3EGJ1639-4702 \\
& & & & &  & & & & & &  PSRJ1637-4642 \\[7 pt]
1AGL J1709-4428  &   17 09 12.6  &  -44 28 44.5  &  343.07    &  -2.64    &    0.20  &  13.8 &  2.18  & 120 $\pm$ 11  &  Pulsar &  PSRJ1709-4429 &  3EGJ1710-4439 \\[7 pt]
1AGL J1736-3235  &   17 36 19.9  &   -32 35 00.8 &   355.85   &   -0.24    &    0.59    &  5.1  &  1.56  & 69 $\pm$ 15  &   Unclassified   &   (C)           &     3EGJ1734-3232 \\[7 pt]  
1AGL J1746-3017  &   17 46 01.5  &   -30 17 23.7 &   358.89   &   -0.78    &    0.68    &  4.4  &  1.56  & 66 $\pm$ 16  &   Unclassified   &   (C)           &     3EGJ1744-3011 \\[7 pt]  

1AGL J180{\bold 1-2317}  &   18 0{\bold 1 22.7} &   -2{\bold 3 17 20.1} &   {\bold 6.66}     &   -0.{\bold 18}    &    0.{\bold 35}    &  {\bold 5.8}  &  1.56  & {\bold 69} $\pm$ 1{\bold 3}  &   {\bold SNR}   &    {\bold W28}           &     3EGJ1800-2338 \\    
& & & & &  & & & & & & {\bold HESSJ1801-233}   \\[7 pt]

{\bold 1AGL J1806-2143}  &   {\bold 18 05 39.5} &   {\bold -21 43 21.2} &   {\bold 8.51}    &   {\bold -0.27}    &   {\bold 0.75}    &  {\bold 4.3}  &  {\bold 1.56}  &  {\bold 54 $\pm$ 13}  &   {\bold Unclassified}   &   {\bold (C) }         &  {\bold PSRJ1803-2137}  \\   
& & & & &  & & & & & &  {\bold W30} \\
& & & & &  & & & & & &  {\bold HESSJ1804-216} \\[7 pt]

{\bold 1AGL J1809-2333}  &   {\bold 18 09 22.8} &   {\bold -23 32 56.3} &  {\bold 7.33}     &  {\bold -1.91}    &   {\bold 0.35}    &  {\bold 5.9}  &  {\bold 1.56}   &  {\bold 53 $\pm$ 10}  &   {\bold Unclassified}  &   {\bold  ---   }        &  {\bold 3EGJ1809-2328}   \\[7 pt] 

{\bold 1AGL J1815-1732}  &   {\bold 18 15 29.7} &   {\bold -17 32 27.1} &  {\bold 13.29}    &  {\bold -0.28}    &   {\bold 0.65}    &  {\bold 4.4}  &  {\bold 1.56}   &  {\bold 52 $\pm$ 13}  &   {\bold Unclassified}  &   {\bold  ---   }        &   {\bold PSRJ1815-1738} \\   
& & & & &  & & & & & &  {\bold HESSJ1813-178}  \\[7 pt]

1AGL J1824-14{\bold 55}  &   18 2{\bold 3 32.5}  &   -14 {\bold 54 41.3} &   1{\bold 6.52}    &   -0.{\bold 74 }   &   {\bold 0.52 }    & {\bold 7.3}  &  1.56  & {\bold 86} $\pm$ 13  &   Unclassified   &   {\bold (C) }            &     3EGJ182{\bold 4-1514} \\   
& & & & &  & & & & & & G16.41−0.55 \\
& & & & &  & & & & & & LS 5039  \\[7 pt]

{\bold 1AGL J1827-1227}  &  {\bold 18 26 57.8}  &  {\bold -12 46 58.6} &  {\bold 18.79}    &  {\bold -0.48}    &   {\bold 0.54}     & {\bold 6.5}  & {\bold 1.56}  &{\bold 79 $\pm$ 13}  &  {\bold Unclassified }  &    {\bold ---}     &   {\bold 3EGJ1826-1302}   \\ 
& & & & &  & & & & & &  {\bold HESSJ1825-137}  \\[7 pt]

1AGL J1836+5923  &   18 36 14.8  &   +59 23 30.4 &   88.84    &   24.99    &    0.17    &  15.6 &  5.52  & 45 $\pm$ 4  &   Unclassified   &   ---           &     3EGJ1835+5918 \\
& & & & &  & & & & & &  BZBJ1841+5906 \\[7 pt]
1AGL J1846+6714  &   18 46 19.6  &   +67 14 17.4 &   97.59    &   25.35    &    0.43    &  7.0  &  5.52  & 20 $\pm$ 4  &   Blazar-FSRQ    &   ---           &   BZQJ1849+6705 \\ 
& & & & &  & & & & & & 4C66.20   \\[7 pt]
1AGL J185{\bold 6+0122}  &   18 5{\bold 5 57.7}  &   +01 {\bold 22 24.5} &   3{\bold 4.67}    &   -0.{\bold 38}    &    0.{\bold 25}    &  1{\bold 1.5} &  3.{\bold 90}  & 1{\bold 23} $\pm$ 1{\bold 2}  &   Unclassified   &   ---           &     3EGJ1856+0114 \\
& & & & &  & & & & & &  {\bold W44} \\
& & & & &  & & & & & &  PSRJ1856+0113 \\[7 pt]

{\bold 1AGL J1901+0429}  &  {\bold 19 01 20.8}  &  {\bold +04 29 38.5} &  {\bold 38.06 }   &  {\bold -0.15 }   &   {\bold 0.58 }   & {\bold 4.4} & {\bold 3.06}  &{\bold 45 $\pm$ 11}  &  {\bold Unclassified }  &  {\bold --- }          &   {\bold PSRJ1901+0435}   \\[7 pt]

1AGL J1908+061{\bold 4}  &   19 08 {\bold 08.4}  &   +06 1{\bold 4 34.5} &   40.3{\bold 9}    &   -0.8{\bold 5}    &    0.4{\bold 5}    &  7.{\bold 7}  &  3.06  & 7{\bold 6} $\pm$ 1{\bold 1}  &   Unclassified   &   ---           &     3EGJ1903+0550 \\
& & & & &  & & & & & &  PSRJ1905+0616 \\[7 pt]
{\bold 1AGL J1923+1404}  &  {\bold 19 22 53.7}  &  {\bold +14 03 45.2} &  {\bold 49.00 }   &  {\bold -0.42 }   &   {\bold 0.64 }   & {\bold 6.6} & {\bold 4.00}  &{\bold 60 $\pm$ 10}  &  {\bold Unclassified}   &   {\bold (C)}           &  {\bold W51}    \\
& & & & &  & & & & & & {\bold PSRJ1921+1419} \\[7 pt] 
1AGL J2021+3652  &   20 21 25.3  &   +36 52 32.6 &   75.28    &   0.07     &    0.19    &  14.2 &  8.31  & 65 $\pm$ 5  &   Pulsar         &   PSRJ2021+3651 &     3EGJ2021+3716 \\[7 pt]	
\hline

\end{tabular}


\normalsize

\begin{minipage}{160mm}
{\bf Table 3.} - Continued \\
\end{minipage}

\end{table*}

\begin{table*}[tbh]


\scriptsize


\begin{tabular}{|lccccccccccc|}

\hline
 & & & &  & & & & & & & \\
\multicolumn{1}{|l}{AGILE name} & 
\multicolumn{1}{c}{RA (J2000.0)} & 
\multicolumn{1}{c}{Dec (J2000.0)} & 
\multicolumn{1}{c}{LII} &
\multicolumn{1}{c}{BII)} &
\multicolumn{1}{c}{$^{a}$Pos. Error (95\%)} & 
\multicolumn{1}{c}{sqrt(TS)} & 
\multicolumn{1}{c}{$^{b}$Mean Ring Exp} &
\multicolumn{1}{c}{$^{c}$Mean Flux \& Error} & 
\multicolumn{1}{c}{Classification} &
\multicolumn{1}{c}{Confirmed Counterp.} &
\multicolumn{1}{c|}{Possible Counterp. }      \\[3 pt]
\multicolumn{1}{|l}{        } & 
\multicolumn{1}{c}{  (hh mm ss) } & 
\multicolumn{1}{c}{  (dd mm ss) } & 
\multicolumn{1}{c}{ (deg) } &
\multicolumn{1}{c}{ (deg) } &
\multicolumn{1}{c}{ (deg)} & 
\multicolumn{1}{c}{   } & 
\multicolumn{1}{c}{  ($\times 10^{8}$ cm$^{2}$ s) } &
\multicolumn{1}{c}{  ($\times 10^{-8}$ ph cm$^{-2}$ s$^{-1}$) } & 
\multicolumn{1}{c}{           } &
\multicolumn{1}{c}{           } &
\multicolumn{1}{c|}{\& Other Names}      \\[3 pt]
\hline
 & & & &  & & & & & & & \\[3 pt]
1AGL J2022+4032  &   20 22 08.5  &   +40 32 13.4 &   78.37    &   2.04     &    0.12    &  23.4 &  8.31  & 120 $\pm$ 7  &   Unclassified   &   ---     &     3EGJ2020+4017 \\
& & & & &  & & & & & &  SNR Gamma Cygni \\[7 pt]
1AGL J2026-0732  &   20 26 30.7  &   -07 32 45.3 &   37.05    &   -24.55   &    0.53    &  6.9  &  3.06  & 39 $\pm$ 7  &   Blazar-FSRQ    &   ---           &     3EGJ2025-0744 \\
& & & & &  & & & & & &  BZQJ2025-0735 \\
& & & & &  & & & & & &  PKS2023-07 \\[7 pt]
1AGL J2032+4102  &   20 32 27.7  &   +41 02 00.0 &   79.91    &   0.74     &    0.41    &  6.8  &  8.31  & 37 $\pm$ 6  &   Unclassified   &   ---           &     3EGJ033+4118  \\ 
& & & & &  & & & & & &  CygX-3 \\[7 pt]
1AGL J2231+6109  &   22 31 07.1  &   +61 09 46.7 &   106.82   &   2.76     &    0.29    &  8.4  &  6.26  & 32 $\pm$ 5  &   Pulsar         &   PSRJ2229+6114 &     3EGJ2227+6122 \\[7 pt]	
1AGL J2254+1602  &   22 54 10.3  &   +16 02 32.6 &   86.09    &   -38.3    &    0.17    &  23.0 &  1.16  & 200 $\pm$ 14  &   Blazar-FSRQ    &   3C454.3       &     3EGJ2254+1601 \\
& & & & &  & & & & & &  BZQJ2253+1608 \\[3 pt]																							  
\hline

\end{tabular}


\normalsize

\begin{minipage}{160mm}
{\bf Table 3.} - Continued \\
$^{a}$ 2D error circle radius at 95\% confidence level, statistical error only. 
The AGILE Team recommends to add linearly a systematic error of $\pm 0.1$ degrees. \\
$^{b}$ mean value of the exposure map relative to the sky area (Ring) used for each source analysis.\\
$^{c}$ E$>$100 {\rm MeV} flux and its $1 \sigma$ statistical error.
The AGILE Team recommends to add to the statistical error a systematic error of $10\%$. \\
\end{minipage}

\end{table*}
\end{landscape}
\twocolumn
